\documentclass{rjparticle}
\usepackage{graphicx}
\usepackage[T1]{fontenc}
\usepackage[latin9]{inputenc}
\usepackage[active]{srcltx}
\usepackage{amsmath}
\usepackage{esint}

\newcommand{\miktex}{\hbox{Mik\kern-.15em\TeX}}

\title{Electrostatic Self-energy of a Charged Particle in the Surroundings
of a Topologically Charged Black Hole in the Brane\\ 
\NoCaseChange{\footnotesize\em \rjpclassfullversion}} 

\author[1]{Alexis Larranaga}
\author[2]{Alejandro Cardenas-Avendano}
\author[3]{Daniel Alexdy Torres}

\affil[1]{National Astronomical Observatory,\\
 National University of Colombia,\\
Ciudad Universitaria, Bogota, Colombia\\
{\em Email}: ealarranaga@unal.edu.co}
\affil[2]{National Astronomical Observatory,\\
National University of Colombia,\\
Ciudad Universitaria, Bogota, Colombia\\
{\em Email}:alcardenasav@unal.edu.co}
\affil[3]{Department of Physics, National University of Colombia,\\
Ciudad Universitaria, Bogota, Colombia\\
{\em Email}:daatorresba@unal.edu.co}


\begin{document}
\maketitle
\begin{abstract}
We determine the self-energy for a point charge held stationary in
a topologically charged black hole spacetime arising from the Randall-Sundrum
II braneworld model, showing that it has two contributions, one of
geometric origin and the other of topological one.
\end{abstract}

\section{Introduction}
The gravitational field modifies the electrostatic interaction of
a charged particle in such a way that the particle experiences a finite
self-force \cite{vilenkin79,Leaute85,Piazzese86,Boisseau96,Smith80,leaute83,zelnikov82}
whose origin comes from the spacetime curvature associated with the
gravitational field. However, even in the absence of curvature it
was shown that a charged point particle \cite{linet86,Smith90} or
a linear charge distribution \cite{Bezerra95} placed at rest may
become subject to a finite repulsive electrostatic self-force (see
also \cite{spinelly00}). In these references, the origin of the force
is the distortion in the particle field caused by the lack of global
flatness of the spacetime of a cosmic string. Therefore, one may conclude
that the modifications of the electrostatic potential comes from two
contributions: one of geometric origin and the other of topological
one. 

In a recent paper \cite{Larranaga2014} our group studied the expression
for the electrostatic potential generated by a point charge held stationary
in the neighborhood of a particular solution of the Randall-Sundrum
braneworld model obtained by Chamblin et. al. \cite{chamblin01} and
revisited by Sheykhi and Wang \cite{sheykhi09-2}, which carry a topological
charge arising from the bulk Weyl tensor. Here we continue this work
by showing that considering this black hole as the background metric,
both kinds of contributions to the self-energy, geometrical and topological,
appear in the self-energy of the charged particle.

\section{The Topologically Charged Black Hole in the Braneworld}

The gravitational field on the brane is described by the Gauss and
Codazzi equations of 5-dimensional gravity \cite{shiromizu00},

\begin{equation}
G_{\mu\nu}=-\Lambda g_{\mu\nu}+8\pi GT_{\mu\nu}+k_{5}^{4}\Pi_{\mu\nu}-E_{\mu\nu},\label{eq:fieldEq}
\end{equation}
where $G_{\mu\nu}$ is the 4-dimensional Einstein tensor and the 5-dimensional
gravity coupling constant is 

\begin{equation}
k_{5}^{4}=\frac{48\pi G}{\tau}
\end{equation}
with $\tau$ the brane tension. $T_{\mu\nu}$ is the stress-energy
tensor of matter confined on the brane and $\Pi_{\mu\nu}$ is a quadratic
tensor in the stress-energy tensor,

\begin{equation}
\Pi_{\mu\nu}=\frac{1}{12}TT_{\mu\nu}-\frac{1}{4}T_{\mu\sigma}T_{\nu}^{\:\:\sigma}+\frac{1}{8}g_{\mu\nu}\left(T_{\alpha\beta}T^{\alpha\beta}-\frac{1}{3}T^{2}\right),
\end{equation}
with $T=T_{\sigma}^{\sigma}$. The tensor $E_{\mu\nu}$ is the projection
of the 5-dimensional bulk Weyl tensor $C_{ABCD}$ on the brane ($E_{\mu\nu}=\delta_{\mu}^{A}\delta_{\nu}^{B}C_{ABCD}n^{A}n^{B}$
with $n^{A}$ the unit normal to the brane). Hence, $E_{\mu\nu}$
encompasses the nonlocal bulk effect and it is traceless, i.e. $E_{\sigma}^{\sigma}=0$.
The 4-dimensional cosmological constant $\Lambda$ is related with
the 5-dimensional cosmological constant $\Lambda_{5}$ by the relation

\begin{equation}
\Lambda=\frac{k_{5}^{2}}{2}\left(\Lambda_{5}+\frac{k_{5}^{2}}{6}\tau^{2}\right).
\end{equation}

In this paper we will consider the Randall-Sundrum scenario in which

\begin{equation}
\Lambda_{5}=-\frac{k_{5}^{2}}{6}\tau^{2}
\end{equation}
implying

\begin{equation}
\Lambda=0.
\end{equation}

Under these considerations, a black hole type solution of the field
equations (\ref{eq:fieldEq}) with $T_{\mu\nu}=0$ is given by the
line element \cite{dadhich2000,chamblin01,sheykhi09-2,Bomer2008}

\begin{equation}
ds^{2}=h\left(r\right)dt^{2}-\frac{dr^{2}}{h\left(r\right)}-r^{2}\left(d\theta^{2}+\sin^{2}\theta d\varphi^{2}\right)\label{eq:lineElement}
\end{equation}
where 
\begin{equation}
h\left(r\right)=1-\frac{2GM}{r}+\frac{\beta}{r^{2}}.
\end{equation}

This metric has a projected Weyl tensor, which transmits the tidal
charge stresses from the bulk to the brane, with components 
\begin{equation}
E_{\: t}^{t}=E_{\: r}^{r}=-E_{\:\theta}^{\theta}=-E_{\:\varphi}^{\varphi}=\frac{\beta}{r^{4}}.
\end{equation}

Thus, the parameter $\beta$ can be interpreted as a tidal charge
associated with the bulk Weyl tensor and hence, there is no restriction
on it to take positive as well as negative values. If $\beta\geq0$
there is a direct analogy to the Reissner-N�rdstrom solution because
the induced metric in the domain wall presents horizons at the radii
(taking $G=1$)

\begin{equation}
r_{\pm}=M\pm\sqrt{M^{2}-\beta}.
\end{equation}

These two horizons lie inside the Schwarzschild radius $r_{s}=2M$,
i.e.

\begin{equation}
0\leq r_{-}\leq r_{+}\leq2M.
\end{equation}

However, there is an upper limit on $\beta$, namely

\begin{equation}
0\leq\beta\leq\beta_{max}=M^{2}
\end{equation}
which corresponds to an extremal back hole case with a degenerate
horizon located at $r_{e}=M$. 

As we have stated before, there is nothing to stop us choosing $\beta$
to be negative, an intriguing new possibility which leads to only
one horizon, 

\begin{equation}
r_{*}=M+\sqrt{M^{2}+\left|\beta\right|}>2M,\label{eq:bnegativehorizon}
\end{equation}
lying outside the corresponding Schwarzschild radius. In this case,
the single horizon has a greater area that its Schwarzschild counterpart.
Thus, one concludes that the effect of the bulk producing a negative
$\beta$ is to strengthen the gravitational field outside the black
hole (obviously it also increase the entropy and decrease the Hawking
temperature). We will show that a negative value of $\beta$ also
produces a decrease in the electrostatic self-energy of a charged
particle located outside the horizon.

Performing the change of the radial coordinate 

\begin{equation}
r=\rho+M+\frac{M^{2}-\beta}{4\rho},
\end{equation}
gives the line element (\ref{eq:lineElement}) in the isotropic coordinate
system $\left(t,\rho,\theta,\varphi\right)$ 

\begin{equation}
ds^{2}=N^{2}\left(\rho\right)dt^{2}-B^{2}\left(\rho\right)\left(d\rho^{2}+\rho^{2}d\Omega^{2}\right)\label{eq:IsotropicLineElement}
\end{equation}
where $N^{2}\left(\rho\right)=\frac{1}{r^{2}}\left[\rho-\frac{M^{2}-\beta}{4\rho}\right]^{2}$and
$B^{2}\left(\rho\right)=\frac{r^{2}}{\rho^{2}}$. Using these coordinates,
the horizons are located at the values $\rho_{\pm}$ defined by the
relation 
\begin{equation}
N\left(\rho_{\pm}\right)=0\label{eq:HorizonsinN}
\end{equation}
and the surface gravity at the horizon is

\begin{equation}
\kappa=\frac{1}{B\left(\rho_{+}\right)}\left.\frac{dN}{d\rho}\right|_{\rho=\rho_{+}}.
\end{equation}

From this expression and assuming that $\kappa B\left(\rho_{+}\right)\neq0$,
we can write 

\begin{equation}
N\left(\rho\right)\propto\kappa B\left(\rho_{+}\right)\left(\rho-\rho_{+}\right)\mbox{ when }\rho\rightarrow\rho_{+}.\label{eq:NneartheHorizon}
\end{equation}

Similar relations for the $\beta<0$ case are obtained by replacing
$r_{+}$ with $r_{*}$ ( or $\rho_{+}$ with $\rho_{*}$ correspondingly).

\section{The Electrostatic Field of a Point Particle}

Copson \cite{copson28} and Linet \cite{linet1976} found the electrostatic
potential in a closed form of a point charge at rest outside the horizon
of a Schwarzschild black hole and that the multipole expansion of
this potential coincides with the one given by Cohen and Wald \cite{cohen71}
and by Hanni and Ruffini \cite{hanni73}. In this section we will
investigate this problem in the background of the topologically charged
black hole (\ref{eq:lineElement}). If the electromagnetic field of
the point particle is assumed to be sufficiently weak so its gravitational
effect is negligible, the Einstein-Maxwell equations reduce to Maxwell's
equations in the curved background (\ref{eq:IsotropicLineElement}).
In the electrostatic case, the spatial components of the potential
vanish, $A^{i}=0$, while the temporal component $A^{0}=\phi$ is
determined by the equation

\begin{equation}
\frac{1}{\sqrt{g}}\partial_{i}\left(\sqrt{g}g^{00}g^{ij}\partial_{j}A_{0}\right)=4\pi J^{0}\label{eq:MaxwellEq}
\end{equation}
where $J^{0}$ is the charge density. When considering a point test
charge $q$ held stationary at $\left(\rho_{0},\theta_{0},\varphi_{0}\right)$,
with $\rho_{0}>\rho_{+}$ in the case $\beta\geq0$ or $\rho_{0}>\rho_{*}$
in the case $\beta<0$, the associated current density $J^{i}$ vanishes
while the charge density $J^{0}$ is given by

\begin{equation}
J^{0}=\frac{q}{\sqrt{g}}\delta\left(\rho-\rho_{0}\right)\delta\left(\theta-\theta_{0}\right)\delta\left(\varphi-\varphi_{0}\right).
\end{equation}

Choosing, without loss of generality, $\theta_{0}=0$, equation (\ref{eq:MaxwellEq})
becomes 

\begin{equation}
\nabla^{2}A_{0}+\frac{N}{B}\frac{\partial}{\partial\rho}\left(\frac{B}{N}\right)\partial_{\rho}A_{0}=-4\pi\frac{q}{\rho^{2}}\frac{N}{B}\delta\left(\rho-\rho_{0}\right)\delta\left(\cos\theta-1\right)\label{eq:EqPhi}
\end{equation}
where $\nabla^{2}$ is the Laplacian operator. The behavior of $A_{0}$
in the neighbourhood of the point $\left(\rho=\rho_{0},\theta=0\right)$
is given by

\begin{equation}
A_{0}\left(\rho,\theta\right)\sim\frac{N\left(\rho_{0}\right)}{B\left(\rho_{0}\right)}\frac{q}{\sqrt{\rho^{2}-2\rho\rho_{0}\cos\theta+\rho_{0}^{2}}}.\label{eq:A0aux1}
\end{equation}

In order to obtain the solution of this equation near the horizon
and in the far region, we expand $A_{0}$ in spherical harmonics,

\begin{equation}
A_{0}\left(\rho,\theta\right)=\sum_{l=0}^{\infty}R_{l}\left(\rho,\rho_{0}\right)P_{l}\left(\cos\theta\right)
\end{equation}
where the radial function satisfies the differential equation

\begin{equation}
\frac{d^{2}R_{l}}{d\rho^{2}}+\left(\frac{2}{\rho}+\frac{1}{B}\frac{dB}{d\rho}-\frac{1}{N}\frac{dN}{d\rho}\right)\frac{dR_{l}}{d\rho}-\frac{l\left(l+1\right)}{\rho^{2}}R_{l}=-q\left(2l+1\right)\frac{N}{B}\frac{\delta\left(\rho-\rho_{0}\right)}{\rho^{2}}.\label{eq:Requation}
\end{equation}

Now we need to define two linearly independent solutions $g_{l}$
and $f_{l}$ with the appropriate boundary solutions, so we will impose
that the electric field derived from the obtained electrostatic potential
should be well behaved at the horizon and at the spatial infinity
of the black hole. For $l=0$, equation (\ref{eq:Requation}) can
be integrated to obtain the condition

\begin{equation}
\frac{dR_{0}\left(\rho\right)}{d\rho}\propto\frac{1}{\rho^{2}}\frac{N\left(\rho\right)}{B\left(\rho\right)}.
\end{equation}

Thus, we consider the functions

\begin{equation}
g_{0}\left(\rho\right)=1
\end{equation}
and

\begin{equation}
f_{0}\left(\rho\right)=a\left(\rho\right)=\int_{\rho}^{\infty}\frac{N\left(\rho'\right)}{\rho'^{2}B\left(\rho'\right)}d\rho'.\label{eq:Definitiona}
\end{equation}

Function $a\left(\rho\right)$ is finite at the horizon and its value
will be denoted $a_{+}=a\left(\rho_{+}\right)$. For $l\neq0$ equation
(\ref{eq:Requation}) has a singular point of regular type at $r=r_{+}$
because from eq. (\ref{eq:NneartheHorizon}) we have

\begin{equation}
\frac{1}{N}\frac{dN}{d\rho}\propto\frac{1}{\rho-\rho_{+}}\mbox{ when }\rho\rightarrow\rho_{+}.
\end{equation}

The roots of the indicial equation relative to this point are $0$
and $2$ and therefore we have a regular solution at the horizon that
we will denote 
\begin{equation}
g_{l}\left(\rho\right)\propto\left(\rho-\rho_{+}\right)^{2}\mbox{ when }\rho\rightarrow\rho_{+},\label{eq:auxCondition}
\end{equation}
and that gives a well behaved electric field at the horizon but is
singular at $\rho\rightarrow\infty$. This fact shows that the black
hole does not have multipole electric moments, except for the monopole.
On the other hand, the regular solution at $\rho\rightarrow\infty$
will be denoted $f_{l}\left(\rho\right)$ and following a similar
argument, it will be singular at the horizon. In conclusion, the electrostatic
potential with the adequate boundary conditions is written

\begin{equation}
A_{0}\left(\rho,\theta\right)=\begin{cases}
qa\left(\rho_{0}\right)+q\sum_{l=1}^{\infty}C_{l}g_{l}\left(\rho\right)f_{l}\left(\rho_{0}\right)P_{l}\left(\cos\theta\right) & \mbox{ for }\rho_{+}<\rho<\rho_{0}\\
qa\left(\rho\right)+q\sum_{l=1}^{\infty}C_{l}g_{l}\left(\rho_{0}\right)f_{l}\left(\rho\right)P_{l}\left(\cos\theta\right) & \mbox{ for }\rho>\rho_{0}
\end{cases}\label{eq:ExpansionA}
\end{equation}
where the constant coefficients $C_{l}$ are determined by equation
(\ref{eq:Requation}).

\section{Electrostatic Self-Energy}

As is well known, the Coulombian part of the electrostatic potential
$A_{0}$ does not yield an electrostatic self-force associated with
the Killing vector $\partial_{0}$. However, the regular part of the
potential at $\rho=\rho_{0}$ and $\theta=0$ defines an electrostatic
self-energy $W_{self}\left(\rho_{0}\right)$. As noted in \cite{Smith80,zelnikov82,leaute83},
this energy can be obtained through the limit

\begin{equation}
\frac{q}{2}\left[A_{0}\left(\rho,\theta\right)-\frac{N\left(\rho_{0}\right)}{B\left(\rho_{0}\right)}\frac{q}{\sqrt{\rho^{2}-2\rho\rho_{0}\cos\theta+\rho_{0}^{2}}}\right]\rightarrow W_{self}\left(\rho_{0}\right)\mbox{ when \ensuremath{\rho\rightarrow\rho_{0}} and \ensuremath{\theta\rightarrow}0. }\label{eq:limit}
\end{equation}

From this expression it is clear that the Coulombian part of the electrostatic
potential, represented by equation (\ref{eq:A0aux1}) does not yield
an electrostatic self-force. Therefore, in order to calculate the
self-energy we will use the result of our previous work \cite{Larranaga2014}
where we have shown that the electrostatic potential of a charged
particle in the background of a topological charged black hole (\ref{eq:lineElement})
satisfies the differential equation (\ref{eq:MaxwellEq}) in isotropic
coordinates with the coefficient 

\begin{equation}
\frac{N\left(\rho\right)}{B\left(\rho\right)}=\frac{\left(1-\frac{\sqrt{M^{2}-\beta}}{2\rho}\right)}{\left(1+\frac{\sqrt{M^{2}-\beta}}{2\rho}\right)^{3}}.
\end{equation}
The solution of the electrostatic solution in isotropic coordinates,
obtained by using the Hadamard method \cite{copson28} and denoted
as $\phi_{C}$, has the same behavior as equation (\ref{eq:A0aux1})
in the neighborhood of the point $\left(\rho=\rho_{0},\theta=0\right)$,

\begin{equation}
\tilde{\phi}\left(\rho,\theta\right)\sim\frac{\left(1-\frac{\sqrt{M^{2}-\beta}}{2\rho_{0}}\right)}{\left(1+\frac{\sqrt{M^{2}-\beta}}{2\rho_{0}}\right)^{3}}\frac{q}{\sqrt{\rho^{2}-2\rho\rho_{0}\cos\theta+\rho_{0}^{2}}}.
\end{equation}

Replacing this condition in equation (\ref{eq:limit}) we obtain the
limit process

\begin{equation}
\frac{q}{2}\left[A_{0}\left(\rho,\theta\right)-\frac{N\left(\rho_{0}\right)}{B\left(\rho_{0}\right)}\frac{\left(1+\frac{\sqrt{M^{2}-\beta}}{2\rho_{0}}\right)^{3}}{\left(1-\frac{\sqrt{M^{2}-\beta}}{2\rho_{0}}\right)}\tilde{\phi}\left(\rho,\theta\right)\right]\rightarrow W_{self}\left(\rho_{0}\right)\label{eq:LimitToPerform}
\end{equation}
as $\rho\rightarrow\rho_{0}$ and $\theta\rightarrow0$. Using Gauss'
theorem it has been shown \cite{Larranaga2014} that $\tilde{\phi}$
describes a secondary charge with value $-\frac{q\sqrt{M^{2}-\beta}}{\rho_{0}\left(1+\frac{\sqrt{M^{2}-\beta}}{2\rho_{0}}\right)^{2}}$
inside the horizon. Hence, the multipole expansion of the potential
$\phi_{C}$ gives 

\begin{equation}
\tilde{\phi}\left(\rho,\theta\right)=\begin{cases}
\frac{q}{\rho_{0}\left(1+\frac{\sqrt{M^{2}-\beta}}{2\rho_{0}}\right)^{2}}\left[1-\frac{\sqrt{M^{2}-\beta}}{\rho\left(1+\frac{m}{2\rho}\right)^{2}}\right]+q\sum_{l=1}^{\infty}\tilde{C}_{l}\tilde{g}_{l}\left(\rho\right)\tilde{f}_{l}\left(\rho_{0}\right)P_{l}\left(\cos\theta\right) & \mbox{ for }\rho_{+}<\rho<\rho_{0}\\
\frac{q}{\rho\left(1+\frac{\sqrt{M^{2}-\beta}}{2\rho}\right)^{2}}\left[1-\frac{\sqrt{M^{2}-\beta}}{\rho_{0}\left(1+\frac{\sqrt{M^{2}-\beta}}{2\rho_{0}}\right)^{2}}\right]+q\sum_{l=1}^{\infty}\tilde{C}_{l}\tilde{g}_{l}\left(\rho_{0}\right)\tilde{f}_{l}\left(\rho\right)P_{l}\left(\cos\theta\right) & \mbox{ for }\rho>\rho_{0}.
\end{cases}\label{eq:ExpansionPhi}
\end{equation}

Replacing the multipole expansions (\ref{eq:ExpansionA}) and (\ref{eq:ExpansionPhi})
in equation (\ref{eq:LimitToPerform}) and setting $\rho=\rho_{0}$
and $\theta=0$ in the infinite series it is possible to evaluate
$W_{self}\left(\rho_{0}\right)$. However, we are interested in the
determination of the electrostatic self-energy on the horizon, so
we also take the limit $\rho_{0}\rightarrow\rho_{+}$. Each term in
the infinite series labeled by $l=1,2,3,...$ contains the polynomials
$g_{l}\left(\rho_{+}\right)$ or $\tilde{g}_{l}\left(\rho_{+}\right)$
which vanish because of property (\ref{eq:auxCondition}). Hence,
only the monopole terms in the multipole expansions contribute to
the limit process, giving the final result as

\begin{equation}
W_{self}=\frac{q^{2}}{2}\left(a_{+}-\kappa\right)\label{eq:SelfEnergy}
\end{equation}
with $a_{+}$ defined in equation (\ref{eq:Definitiona}) and the
surface gravity $\kappa$ appearing in the limit process through

\begin{equation}
\kappa=\lim_{\rho_{0}\rightarrow\rho_{+}}\frac{N\left(\rho_{0}\right)}{\rho_{+}B\left(\rho_{+}\right)\left(1-\frac{\sqrt{M^{2}-\beta}}{2\rho_{0}}\right)}
\end{equation}
where we used equation (\ref{eq:NneartheHorizon}). Obviously, the
self-energy given in (\ref{eq:SelfEnergy}) is independent of the
choice of the radial coordinate. In terms of the radial coordinate
$r$ in which the line element of the topologically charged black
hole is given by equation (\ref{eq:lineElement}), it is straightforward
to calculate

\begin{equation}
a_{+}=\int_{r_{+}}^{\infty}\frac{dr}{r^{2}}=\frac{1}{r_{+}}
\end{equation}
and 
\begin{equation}
\kappa=\frac{\sqrt{M^{2}-\beta}}{r_{+}^{2}},
\end{equation}
and therefore the self-energy is simply

\begin{equation}
W_{self}=\frac{q^{2}}{2}\frac{M}{r_{+}^{2}}.
\end{equation}

A similar analysis can be performed in order to consider the $\beta<0$
black hole case, obtaining the final result

\begin{equation}
W_{self}=\frac{q^{2}}{2}\frac{M}{r_{*}^{2}}
\end{equation}
where the horizon radius $r_{*}$ is given by equation (\ref{eq:bnegativehorizon}).

\begin{figure}
\begin{centering}
\includegraphics[scale=0.3]{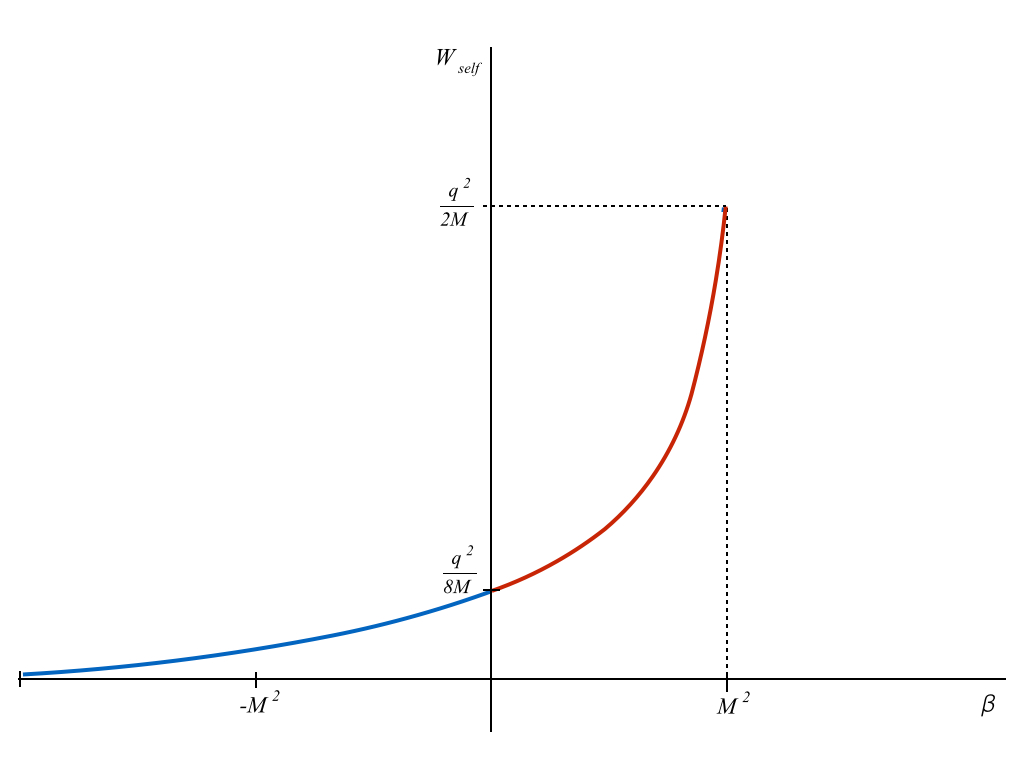}
\par\end{centering}

\caption{Self-energy of a particle with electric charge $q$ in the background
of the topologically charged black hole as function of the topological
charge $\beta$. The red section gives the behavior for $0<\beta<M^{2}$
while the blue section shows the self-energy when $\beta<0$.}
\end{figure}

Figure 1 shows the complete behavior of the self-energy of the particle
with electric charge $q$ in the background of the topologically charged
black hole. The red section gives the behavior for a topological charge
in the range $0<\beta<M^{2}$ while the blue section shows the self-energy
when $\beta<0$. Note that the extremal black hole, characterized
by the condition $\beta=M^{2}$, gives a self-energy $W_{self}=\frac{q^{2}}{2M}$
which is the same value obtained for the extreme Reissner-Nordstr�m
black hole \cite{Linet2000,leaute76}, due to the similarities between
both metrics. Similarly, the self-energy for the Schwarschild metric
corresponds to the point $\beta=0$ giving $W_{self}=\frac{q^{2}}{8M}$
\cite{Smith80,leaute83,zelnikov82}. Finally, it is interesting to
note that the self-energy of the particle goes to zero as the bulk
geometry is such that $\beta\rightarrow-\infty$, a behavior that
never appears in the Reissner-Nordstr�m case. This fact shows that
the complete curve in Figure 1 represents a self-energy which has
two kinds of contributions, one of geometric origin (through the mass
of the black hole) and the other of topological one (appearing in
our analysis through the bulk Weyl tensor). 

\begin{acknowledgement}
This work was supported by the Universidad Nacional de Colombia. Hermes
Project Code 18140.
\end{acknowledgement}

\end{document}